\documentclass[usenatbib]{mn2e}
\usepackage{epsfig}
\usepackage{subfigure}
\usepackage{amssymb}
\usepackage{color}
\usepackage{comment}
\usepackage{float}
\usepackage[T1]{fontenc}
\usepackage{aecompl}

\newcommand{\bn}{\begin{enumerate}}
\newcommand{\en}{\end{enumerate}}
\newcommand{\bi}{\begin{itemize}}
\newcommand{\ei}{\end{itemize}}

\newcommand{\Msun}{\rm M_{\odot}}
\newcommand{\Mach}{\mathcal M}

\def\gtorder{\mathrel{\raise.3ex\hbox{$>$}\mkern-14mu
    \lower0.6ex\hbox{$\sim$}}}
\def\ltorder{\mathrel{\raise.3ex\hbox{$<$}\mkern-14mu
    \lower0.6ex\hbox{$\sim$}}}

\newcommand{\apj}{ApJ}
\newcommand{\aap}{A\&A}
\newcommand{\apjl}{ApJL}
\newcommand{\apjs}{ApJS}
\newcommand{\mnras}{MNRAS}
\newcommand{\aj}{AJ}
\newcommand{\araa}{ARA\&A}

\newcommand{\nat}{Nature}

\newcommand{\na}{New Astronomy}
\newcommand{\prd}{Phys.~Rev.~D}

\title[Supermassive Black Hole Formation at High Redshifts]
{Supermassive Black Hole Formation at High Redshifts via Direct Collapse in a Cosmological Context}

\author[Choi, Shlosman and Begelman]
{Jun-Hwan Choi$^{1}$\thanks{E-mail: jhchoi@astro.as.utexas.edu}, 
 Isaac Shlosman$^{2,3}$\thanks{E-mail: shlosman@pa.uky.edu}, 
     Mitchell C. Begelman$^{4,5}$
\\
$^{1}$ Department of Astronomy, University of Texas, Austin,  TX 78712-1205 , USA \\
$^{2}$ Department of Physics \& Astronomy, University of Kentucky, Lexington, KY 40506-0055, USA\\
$^{3}$ Theoretical Astrophysics, Department of Earth \& Space Science, Osaka University, 
     Osaka 560-0043, Japan\\
$^{4}$ JILA, University of Colorado and National Institute of Standards and Technology, 
     440 UCB, Boulder, CO 80309-0440, USA\\
$^{5}$ Department of Astrophysical and Planetary Sciences, 391UCB, University of Colorado, 
     Boulder, CO 80309-0391, USA\\ 
}

\begin{document}

\date{Accepted ?; Received ??; in original form ???}

\pagerange{\pageref{firstpage}--\pageref{lastpage}} \pubyear{2014}

\maketitle

\label{firstpage}

\begin{abstract}
We study the early stage of the formation of seed supermassive black holes via 
direct collapse in dark matter (DM) halos, in the cosmological context. 
We perform high-resolution zoom-in simulations of such collapse at high-$z$.
Using the adaptive mesh refinement code ENZO, we resolve the formation and growth 
of a DM halo, until its virial temperature reaches $\sim 10^4$\,K, atomic cooling 
turns on, and collapse ensues. We demonstrate that direct collapse proceeds 
in two stages, although they are not well separated.  The first stage is triggered 
by the onset of atomic cooling, and leads to rapidly increasing accretion rate with 
radius, from $\dot M\sim 0.1\,M_\odot\,{\rm yr^{-1}}$ at the halo virial radius to 
few $M_\odot \,{\rm yr^{-1}}$, around the scale radius $R_{\rm s}\sim 30$\,pc of 
the NFW DM density profile. The second stage of the collapse commences 
when the gas density takes precedence over the DM density. This is associated with 
the gas decoupling from the DM gravitational potential. The ensuing collapse 
approximates that of an isothermal sphere with  $\dot M (r)\sim $\,const. 
We confirm that the gas loses its angular momentum through non-axisymmetric perturbations 
and gravitational torques, to overcome the centrifugal barrier. During the course of the collapse, this 
angular momentum transfer process happens on nearly all spatial scales, and the angular momentum 
vector of the gas varies with position and time.  Collapsing gas also exhibits 
supersonic turbulent motions which suppress gas fragmentation, and are characterized 
by density PDF consisting of a lognormal part and a high-density power law tail.

\end{abstract}


\section{Introduction}
\label{sec:intro}

Recent observations of quasars at $z\gtorder 6$ \citep[e.g.,][]{Fan.etal:03,Mortlock.etal:11} suggest that 
some supermassive black holes (SMBHs) quickly achieve masses of $M_{\rm \bullet}\gtorder 10^9 \Msun$, 
even before the Universe is a billion years old. The origin of these SMBHs, located in galactic 
centers and serving as energy sources for active galactic nuclei (AGN), is one of the unsolved mysteries of 
contemporary astrophysics. 
The observed correlations between SMBH masses and host galaxy properties, e.g., bulge masses and stellar 
dispersion velocities
\citep[e.g.,][]{Gebhardt.etal:00,Ferrarese.Merritt:00,Tremaine.etal:02}, point to some kind of SMBH-galaxy 
co-evolution. Therefore, the formation of SMBHs
should be considered alongside overall structure formation in the universe. 

In addition to primordial SMBH seeds \citep[e.g.,][]{Carr.etal:10}, several scenarios for early SMBH 
formation exist, such as growth from Population III remnants 
\citep[e.g.,][]{Haiman.Loeb:01,Abel.etal:02,Bromm.Larson:04,Yoo.Miralda-Escude:04,Volonteri.Rees:06,Li.etal:07,
Pelupessy.etal:07,Tanaka.Haiman:09}, and collapse of stellar clusters \citep[e.g.,][]{Devecchi.Volonteri:09,
Lupi.etal:14}. Population III progenitors, originally estimated to be as massive as $\sim 1000\,M_\odot$, 
have been recently downsized by a factor of 10, due to radiation feedback-limited mass and fragmentation (e.g.,
\citealt{Turk.etal:09,Hosokawa.etal:11,Wise.etal:12,Hirano.etal:14}, see also \citealt{Bromm:13} for review).
To explain the detected high-$z$ quasars, Pop III SMBH seeds are required to grow from $\sim 10\,M_\odot$
to $\gtorder 10^9\,M_\odot$ in less than $\sim 7\times 10^8$\,yrs, which is uncomfortably close to 
the age of the Universe at this redshift. Is it plausible that super-Eddington accretion rates persist for 
$\sim 1$\,Gyr? Of course, Pop\,III  remnants can 
lead to less massive SMBHs. On the other hand, models involving relativistic instabilities in stellar clusters 
must explain the origin of these clusters in the first place, and a substantial metal enrichment at these 
high redshifts.   

Models, where early SMBHs seeds formed via direct collapse of gas into dark matter (DM) halos 
at $z\sim 10-20$, provide an attractive alternative 
\citep[e.g.,][]{Oh.Haman:02,Bromm.Loeb:03,Haehnelt.Rees:93,Volonteri.Rees:05,Begelman.etal:06,Wise.etal:08,
Regan.Haehnelt:09,Begelman.Shlosman:09,Milosavljevic.etal:09,Mayer.etal:10,Schleicher.etal:10,
Hosokawa.etal:11,Johnson.etal:11,Prieto.etal:13,Choi.etal:13,Latif.etal:13a,Latif.etal:13b}. Gas collapse 
is triggered by atomic gas cooling, 
and occurs when the halo virial temperature surpasses $T_{\rm vir}\sim 10^4 K$.  Several recent studies have 
dealt with the halo population hosting massive SMBH seeds 
\citep{Prieto.etal:13,Agarwal.etal:13,Agarwal.etal:14}. 

Within the direct collapse framework, two alternative pathways have been proposed. First, a massive 
central object --- a supermassive star (SMS) --- forms at the center of the DM halo and is 
powered by a combination of core nuclear burning and Kelvin-Helmholtz 
contraction \citep[e.g.,][]{Begelman.etal:06,Begelman.etal:08,Begelman:10}. Following collapse of the
stellar core and formation of the SMBH seed of $\sim 10 - {\rm few}\times 100\,\Msun$, depending on the 
angular momentum distribution \citep{Begelman:10}, its convective envelope is powered by 
super-Eddington accretion onto this seed --- such a configuration has been termed a `quasistar.' 
The seed SMBH can grow to $\sim 10^{5-6}\,\Msun$ in less than few Myrs.
An important ingredient in the formation of the SMS is the very high gas accretion rate, $\gtorder 
0.1\,\Msun\,{\rm yr^{-1}}$ \citep{Begelman.etal:06,Hosokawa.etal:2013,Schleicher.etal:13}. A basic
ingredient of this model is trapping of the escaping energy and momentum within the SMS and the quasistar. 

Recent work by \citet{Becerra.etal:14} has used cosmological simulations with the AREPO moving mesh code 
\citep{Springel:10}. Probably the most important assumption made was that the cooling rate
of the collapsing gas has been artificially and exponentially suppressed at densities above 
$10^{16}\,{\rm cm^{-3}}$. 
This led to a sharp increase in the gas temperature above these densities and the truncation of the
gravitational collapse. As a result, the formation of the SMS has been supplemented by subsequent
accretion from the surrounding disk, and by the disk fragmentation into low-mass stars.  

According to the second pathway, gravitational collapse can retain a disky character even at the innermost 
scales, and, in tandem with the existence of a preferred channel for momentum and energy release (e.g., 
jets or winds), can bypass the SMS and quasistar stages, and the associated thermonuclear reactions  
\citep[e.g.,][]{Begelman.Shlosman:09,Choi.etal:13}. 

Two caveats to the direct collapse scenario have been singled out because of their importance 
\citep[e.g.,][]{Begelman.Shlosman:09}. First, the angular momentum barrier, in principle, can terminate the 
collapse well before it reaches $\sim 1-10$\,AU scale. Second, the gas could fragment, depleting the 
accretion stream by forming clumps and ultimately stars, and disturbing the accretion pattern. 
\citet{Choi.etal:13} have performed a baseline study of direct collapse under idealized conditions, when 
the DM halo is isolated and the gas is in rotational equilibrium with a cosmological spin
$\lambda\sim 0.05$. They confirmed that gas 
within slowly tumbling DM halos of $M_{\rm vir}\sim 10^8\,\Msun$ and $R_{\rm vir}\sim 1\,$kpc can 
collapse to $\sim 10$\,AU scale. The collapsing flow overcomes the centrifugal barrier, losing its angular 
momentum by breaking axial symmetry and forming nested gaseous bars. Ultimately, its angular momentum is 
removed by gravitational torques from
the gas and the DM. In the collapse phase, virial 
supersonic turbulence develops and fragmentation is damped. The gas accretion rate exceeds 
$\sim 1\Msun\,{\rm yr^{-1}}$ at various spatial scales, and allows for the formation of SMBH seeds at high 
redshifts, at least in principle.

In this work, we explore direct collapse in a full cosmological setup, and study 
the physical processes associated with its early stages within DM halos with virial 
temperatures of  
$\sim 10^4\,$K. We apply self-consistent zoom-in cosmological simulations. Our goal is to compare the 
gravitational collapse model in cosmological and isolated halos and to target the angular momentum and 
fragmentation problems.  
In Section~\ref{sec:method}, we explain the numerical details and the initial 
conditions. Sections~\ref{sec:prop} and \ref{sec:dyn} provide the results. The general DM halo properties 
are discussed in Section~\ref{sec:prop}, and the dynamical aspects of the gas collapse are analyzed in 
Section~\ref{sec:dyn}. This is followed by discussion and conclusions.

\section{Numerical technique}
\label{sec:method}

\subsection{Numerical Resolution}
\label{sec:res}

\begin{figure*}
\centerline{\includegraphics[width=0.93\textwidth,angle=0] {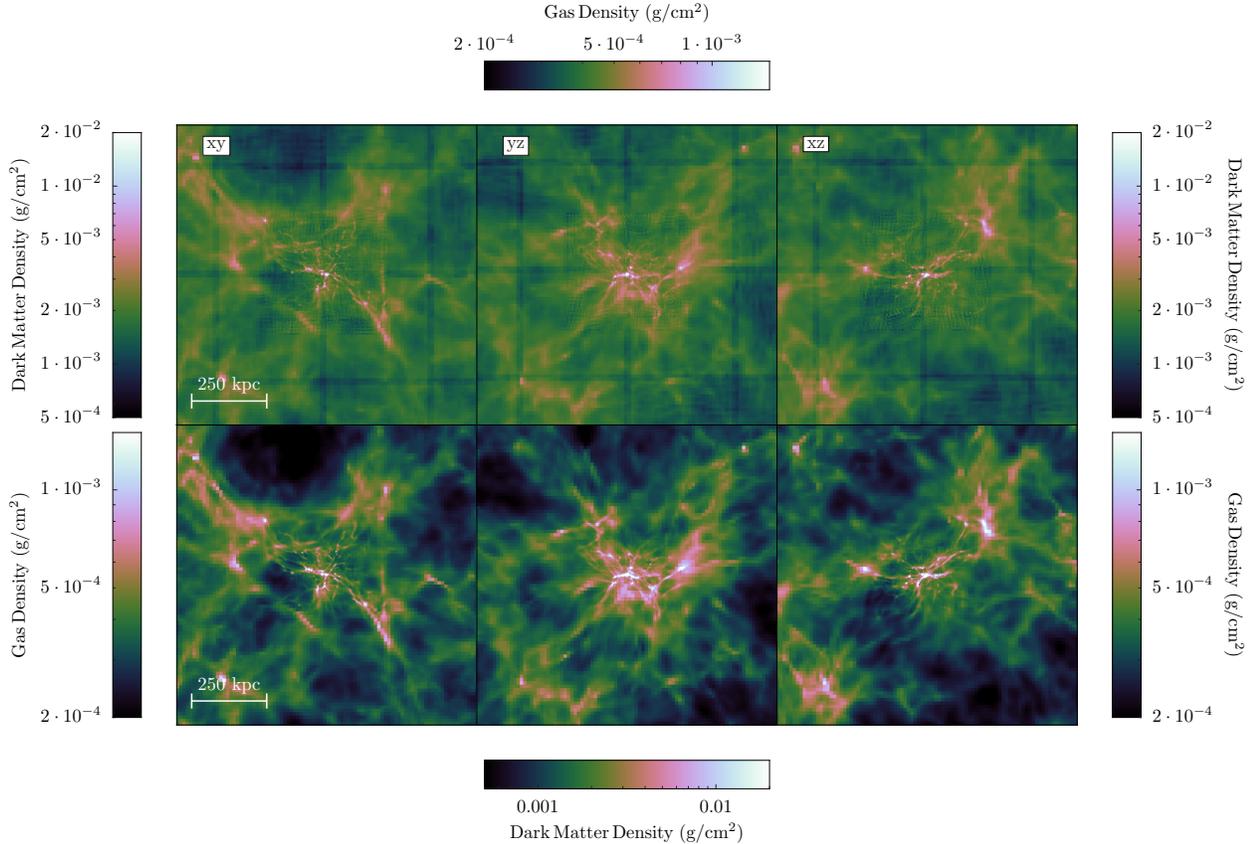}}
\caption{Density-weighted DM and gas density projections of the large-scale  environment of the
targeted DM halo, on scales $\sim 250h^{-1}$\,kpc (comoving), at the end of the simulation, $t\sim 360.13$\,Myr
($z\sim 12$). 
}
\label{fig:ZoomIn0}
\end{figure*}

\begin{figure*}
\centerline{\includegraphics[width=0.93\textwidth,angle=0] {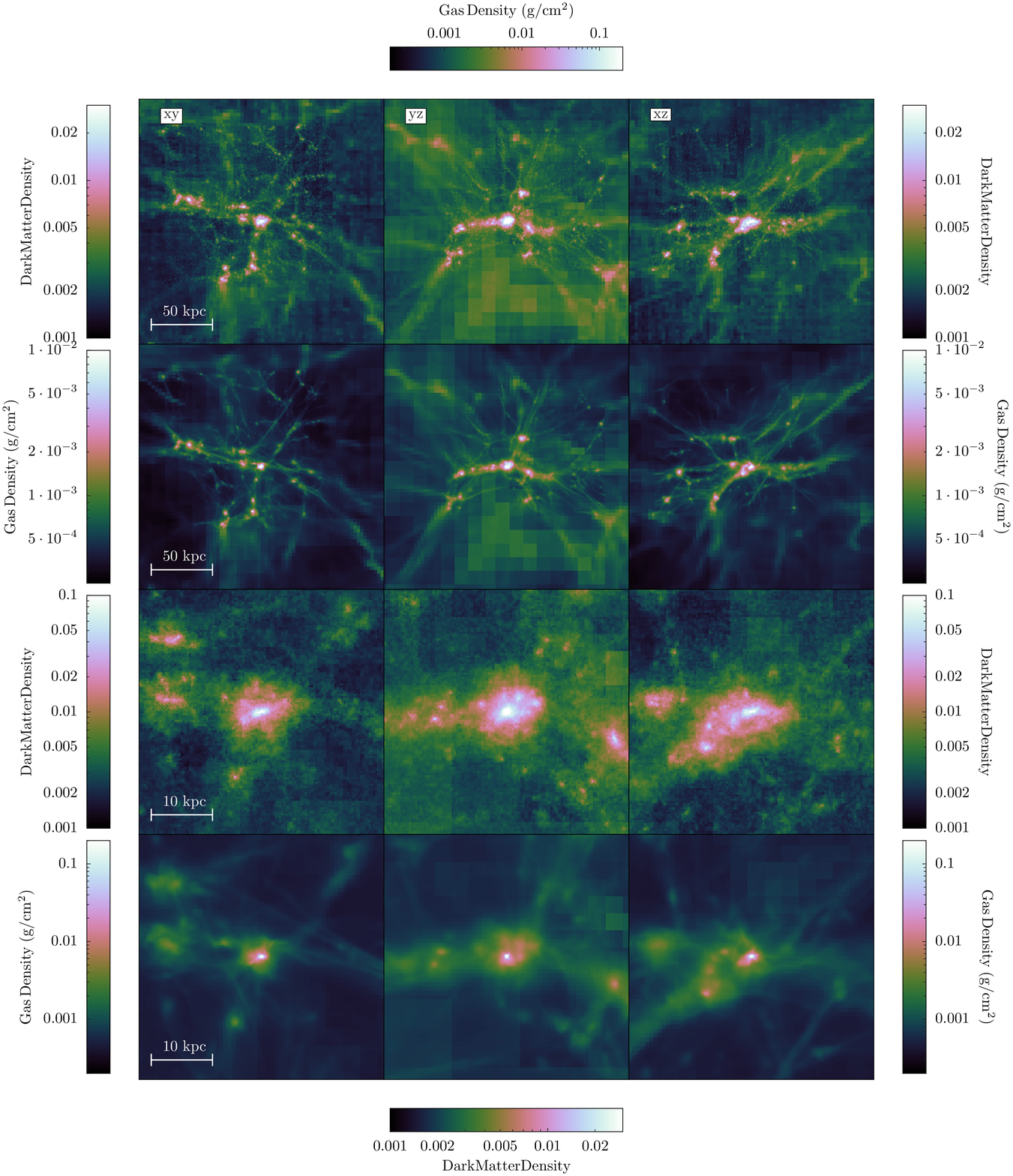}}
\caption{Density-weighted DM and gas density projections of the environment of the
targeted DM halo, on scales $\sim 10 - 50h^{-1}$\,kpc (comoving), at the end of the simulation,
$t\sim 360.13$\,Myr ($z\sim 12$). 
}
\label{fig:ZoomIn}
\end{figure*}

In this study, we use the Eulerian adaptive mesh refinement (AMR) code ENZO-2.3, which has been tested 
extensively and is publicly available \citep{Bryan.Norman:97,Norman.Bryan:99,ENZO:14}. ENZO uses a 
particle-mesh $N$-body method to calculate the gravitational dynamics, including collisionless DM particles, 
and 
a second-order piecewise parabolic method \citep[PPM,][]{Bryan.etal:95} to solve hydrodynamics. The 
structured AMR used in ENZO places no fundamental restrictions on the number of rectangular grids used to 
cover some region 
of space at a given level of refinement, or on the number of levels of refinement \citep{Berger.Colella:89}. 
A region of the simulation grid is refined by a factor of 2 in lengthscale, if either the gas or DM density 
become greater than $\rho_{\rm 0,gas,dm} N^l$, where $\rho_{\rm 0,gas,dm}$ is the cosmic mean density for the 
gas or DM respectively, $N = 2$ is the refinement factor, and $l=35$ is the maximal AMR refinement level. This 
refinement corresponds to a spatial resolution of $\sim 0.005$\,AU.

The \citet{Truelove.etal:97} requirement for resolution of the Jeans length, i.e., at least 4 cells, has been 
verified. Recently, several numerical experiments \citep{Sur.etal:10,Federrath.etal:11,Turk.etal:12,Latif.etal:13a} 
have concluded that even finer refinement for a given Jeans length is required to properly resolve the 
turbulent motions. Accordingly, we have resolved the Jeans length with 32 cells in the simulations.

\subsection{Zoom-in simulations}
\label{sec:zoom}

\begin{figure*}
\centerline{
 \includegraphics[width=0.55\textwidth,angle=0] {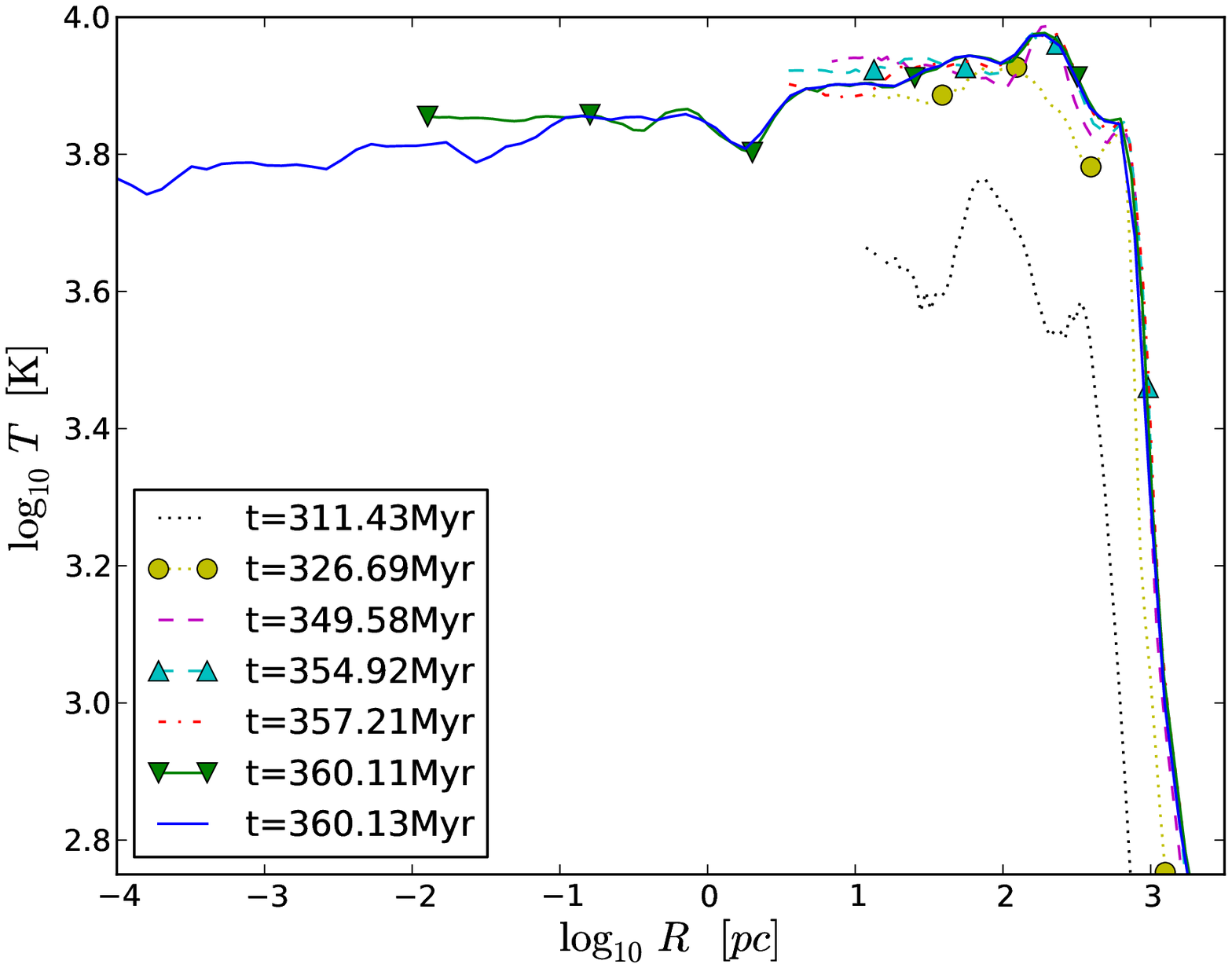}
   \includegraphics[width=0.55\textwidth,angle=0] {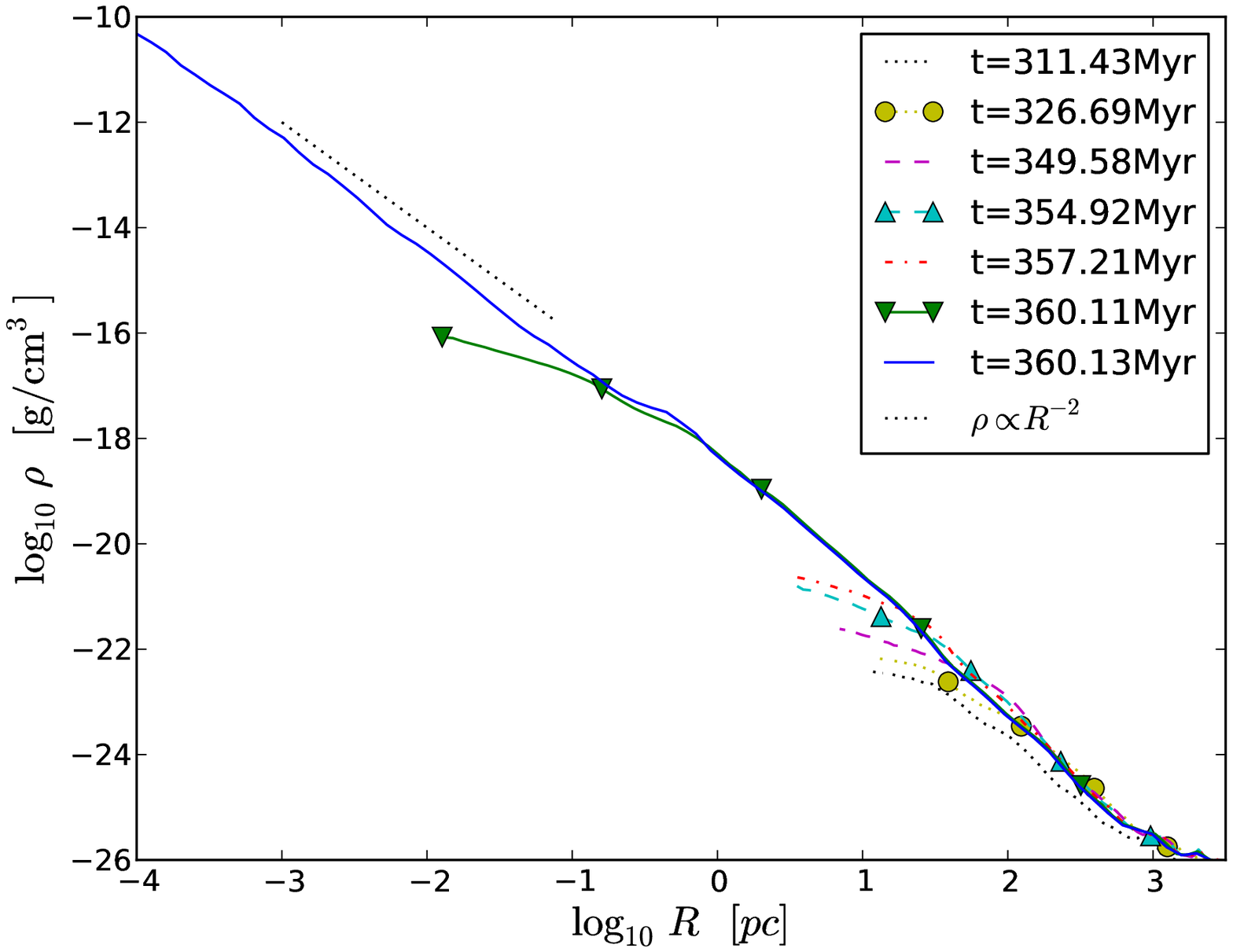}
}
\caption{Evolution of spherically-averaged gas temperature (left) and density (right) profiles.
The time in the legend shows the age of the universe. 
The initial profile corresponds to the time when the halo virial mass is 
sufficient to trigger the atomic cooling in the gas.
The last profile corresponds to time when the gas collapse has reached down to $\sim 10$\,AU scale.
The $x$-axis provides the distance from the densest cell, and is measured in physical units.
The density profiles confirm that the halo gas experiences nearly isothermal central runaway 
collapse.
}
\label{fig:Profile}
\end{figure*}

ENZO follows the non-equilibrium evolution of six species: $\rm H, \; H^{+}, \; He, \; He^{+},  \; He^{++}$, and 
$e^{-}$ \citep{Abell.etal:97,Anninos.etal:97} in a gas with a primordial composition. It calculates radiative 
heating and cooling following atomic line excitation, recombination, collisional excitation and free-free 
transitions.  
Radiative losses from atomic cooling are computed in the optically-thin limit. As discussed in \citet[][and 
references therein]{Choi.etal:13}, several physical processes have been suggested to 
prevent $\rm H_2$ formation, 
such as a very stong Lyman-Werner background radiation \citep[e.g.][]{Dijkstra.etal:08,Ahn.etal:09},
Ly-$\alpha$ photon trapping \citep[e.g.][]{Spaans.Silk:06,Choi.etal:13}, and the collisional dissociation in the 
shocked gas \citep[e.g.,][]{Inayoshi.Omukai:12}.
In this work, we neglect the H$_2$ formation and destruction processes altogether, as well as exclude the chemistry 
and cooling related to $\rm H_2$.
Understanding of $\rm H_2$ suppression mechanism can be important
in estimating the population of the high-$z$ SMBH seeds \citep[e.g.][]{Agarwal.etal:14}.

In this study, we are interested in the detailed dynamical evolution of the collapsing gas within a DM halo in 
the fully cosmological environment and subject to atomic cooling. To satisfy the resolution requirement, we 
use the MUSIC code \citep{Hahn.Abel:11} to generate the cosmological zoom-in initial conditions (ICs). MUSIC 
uses a real-space 
convolution approach in conjunction with an adaptive multi-grid Poisson solver to generate highly accurate nested 
density, particle displacement, and velocity fields suitable for multi-scale zoom-in simulations of structure 
formation in the universe. Generating a set of ``zoom-in'' ICs is a two-step process.
First, we generate $1h^{-1}\,{\rm Mpc}$ comoving 128$^3$ DM-only ICs for the pathfinder simulation and run it 
without AMR until $z=10$. Using the HOP group finder \citep{Eisenstein.Hut:98}, we select an appropriate DM 
halo, whose mass 
is $\sim 10^8h^{-1}\,\Msun$ at $z=10$. Second, we generate $1\,{\rm Mpc}/h$ ICs with 
512$^3$ resolution in DM and gas. Since we use the same random seeds of these ICs as the ICs at the first step, 
the phases of both ICs are identical.
The zoom-in region is centered on the selected halo position and is set to be large enough to cover the initial 
positions of all selected halo particles (see Figures\,\ref{fig:ZoomIn0} and \ref{fig:ZoomIn}).  
We set the DM particle smoothing length at $0.24 h^{-1}\,{\rm pc}$ in comoving coordinates.
The ICs are generated using WMAP5 cosmology:
$\Omega_{\rm \Lambda} = 0.721$, $\Omega_{\rm m} = 0.279$, $\Omega_{\rm b} = 0.0445$, $h=0.701$, $\sigma_8 = 0.807$, 
and $n_{\rm s} = 0.961$. In the following we use $R$ for spherical coordinates and $r$ for cylindrical 
ones.

\section{Results}

The computational box contains a number of DM halos whose virial temperatures exceed $10^4$\,K at $z=10$.
We follow a representative DM halo of our choice down to $z\sim 12$ and observe its growth from mergers and 
accretion. Around $t\sim 350$\,Myr, the DM halo has reached the virial mass and radius of $M_{\rm h}\sim 2\times 
10^7h^{-1}\Msun$ and $R_{\rm vir}\sim 10h^{-1}{\rm kpc}$ in {\it comoving} coordinates, and has acquired cosmological 
spin $\lambda\sim 0.03$. The DM density profile is well approximated by the NFW profile \citep{NFW:97} with the characteristic radius
of $R_{\rm s}\sim 30$\,pc (in physical coordinates), beyond which the DM density profile steepens gradually 
to $\sim -3$.  The DM halo concentration parameter is $c\sim 25$. This halo and its environment are shown on various 
(comoving) spatial scales, from $250h^{-1}$\,kpc 
down to $10h^{-1}$\,kpc, in Figures\,\ref{fig:ZoomIn0} and \ref{fig:ZoomIn}, at the end of the simulation, 
$t\sim 360.13$\,Myr.

The filamentary structure of DM is evident on all scales in Figures\,\ref{fig:ZoomIn0} and \ref{fig:ZoomIn}. 
The gas distribution
follows that of the DM. The targeted halo becomes dominant on the smallest scales shown (bottom 
frames of Fig.\,\ref{fig:ZoomIn}). It is
connected to the DM web via three filaments whose width is nearly comparable to the halo virial diameter. 

The most important feature of the target DM halo is its triaxiality, i.e., its three major axes all differ 
from each other, in all three projection planes, $xy, yz$, and $xz$. This is not suprising, as DM halos
are universally triaxial as they form in numerical simulations \citep[e.g.,][]{Allgood.etal:06} before dissipative
processes axisymmetrize them in the subsequent evolution \citep[e.g.,][]{Berentzen:06,Shlosman:07}. Moreover,
the rotation of the halo figure, i.e., its rate of tumbling, is extremely slow, and seems to be the general property 
of DM halos, as pointed out by \citet[][]{Romano-Diaz.etal:09}.

\subsection{Properties of the collapsing gas in DM halo}
\label{sec:prop}

Since we are interested in details
of the dynamical process within this halo, we switch to physical units.
At the early stage of halo growth, the gas mostly follows the DM assembly. This means that it is 
accumulating within the growing halo, and, during the quiescent time periods of no major mergers, the 
gas is largely hydrostatic with a small degree of the rotational support. Of course `hydrostatic' has a 
very approximate meaning here, as the gas joins the halo partly via penetrating filaments which result 
in large-scale motions within the halo, i.e., the streamers. 
The halo gas stops increasing its  
temperature and starts to cool via atomic cooling, when the halo becomes massive enough and its virial 
temperature has reached 
$\sim 10^4$K. 
Note that we ignore the H$_2$ cooling and the associated chemistry, and implement only atomic cooling (see 
Section~\ref{sec:zoom}).

Cooling allows the gas to be driven into the gravitational 
potential minimum. This is demonstrated in Figure~\ref{fig:Profile} which shows the evolution of gas 
temperature and density profiles within the halo. When $T_{\rm vir}$ reaches $10^4$K, atomic cooling 
becomes important and collapse is triggered.
Low-$T$ gas continues to be accreted from outside $R_{\rm vir}$ along the filaments as 
well as smooth
accretion from arbitrary directions. The smooth accretion experiences a shock at $R\sim 800$\,pc,
which virializes it. The gas being accreted along the filaments virializes about a decade deeper in $R$.
The collapsing gas roughly maintains isothermality at the cooling floor, with a small radial decline in $T$,
as the heating appears to be inefficient. The gas collapse leads to the establishment 
of an isothermal density profile $\rho\propto R^{-2}$ inside $R_{\rm s}$. Finally, the collapse reaches 
$\sim 10$\,AU scale.

\begin{figure}
  \includegraphics[width=1.1\columnwidth,angle=0] {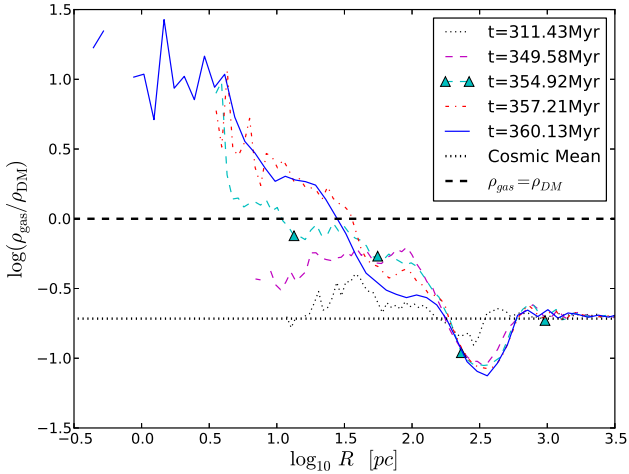}     
\caption{Spherically-averaged gas-to-DM density ratio profiles.
The time in the legend gives the age of the universe.
Initially, the DM and the gas density profiles exhibit very similar shapes reflecting the cosmic mean where
$\rho_{\rm gas}/\rho_{\rm DM}\sim 0.17$. The cooling allows the gas to collapse, increasing the ratio, and to 
reach the DM density at $R_{\rm s}\sim 30$\,pc. This radius movies only slightly inward with time. The outer
region, $r\sim 200-500$\,pc becomes gas deficient (compared to the cosmological mean) as the gas inflow across
$r_{\rm vir}$ cannot replenish the collapsing gas on a short timescale.
}
\label{fig:CompDensity}
\end{figure}

The collapse clearly proceeds from outside in (Figure~\ref{fig:Profile}, right frame). 
Figure~\ref{fig:CompDensity}, which shows the gas-to-DM density ratio profiles at various times, reveals 
the critical detail of this collapse. At $R\gtorder 100$\,pc,  the gas closely follows
the DM density profile, and the baryon-to-DM ratio is at about its cosmic average.
Once the gas temperature is reduced below the virial temperature, this ratio increases in the 
inner halo, and eventually exceeds unity. The dotted-gashed line at $\sim 355$\,Myr corresponds 
to the time when the inner gas density, inside $\sim 5-10$\,pc, exceeds that of the DM. After this 
stage, the inner gas 
density rapidly increases and establishes the isothermal density profile  --- as the second stage of the 
collapse develops, with a very high inflow rate. The gas essentially decouples from the background DM
potential. Similar behavior has been found in the evolution of
isolated models \citep{Choi.etal:13}.

\begin{figure}
\centerline{
  \includegraphics[width=1.1\columnwidth,angle=0] {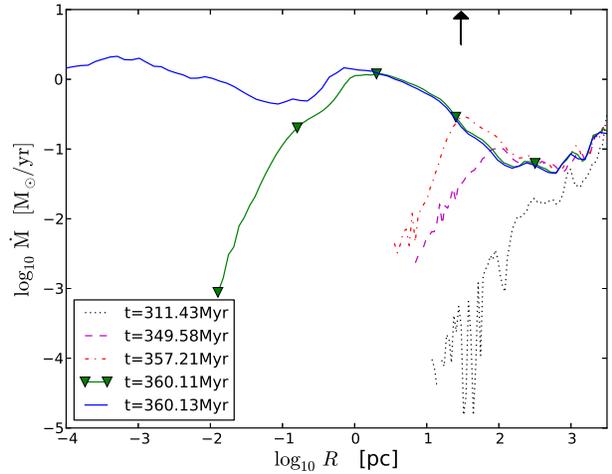}
}
\caption{
Evolution of the spherically-averaged gas inflow rate.
The $R$-axis is the distance from the densest cell measured in physical units.
The accretion rate, $\dot{M}$,  increases from the virial radius to the inner $\sim 10$\,pc scale, which reflects 
the increase in the free-fall velocity within the NFW DM potential.
$\dot{M}$ saturates at around $\sim {\rm few}\,\Msun\,{\rm yr^{-1}}$ in the inner halo, which extends from 
$r\sim 10$\,pc (i.e., just inside $\sim R_{\rm s}$) all the way inward.
The location of the $R_{\rm s}$ is indicated with the black arrow.
}
\label{fig:Mdot}
\end{figure}

Figure~\ref{fig:Mdot} shows the development of the gas inflow rate. From $R_{\rm vir}$
and down to a $\sim 10$\,pc scale (just inside $R_{\rm s}$), it increases from $\dot{M}\sim 
0.03\,\Msun\,{\rm yr^{-1}}$ to 
$\dot{M}\sim {\rm few}\,\Msun\,{\rm yr^{-1}}$, i.e., by nearly two orders of magnitude. Inside $R_{\rm s}$, 
$\dot{M}$ becomes approximately constant with radius. The dramatic increase
in  $\dot{M}$ is a reflection of the steep DM density distribution outside $R_{\rm s}$ and associated gas
distribution, with a logarithmic
slope varying from $\sim -3$ down to $\sim -2$ at $R_{\rm s}$. As long as this DM profile persists,
the mass flux into the inner halo ($\ltorder R_{\rm s}$) will not change: the DM distribution creates  
a kind of bottleneck for the gas supply rate. 
Only growth in the DM virial mass will allow a higher $\dot{M}$ outside $R_{\rm s}$, but the growth of the 
DM halo proceeds on a timescale much 
longer then the gravitational collapse at its center. An interesting corollary of this effect is that after
the halo gas has collapsed to the center, the DM halo will be largely depopulated of gas. 
 
\begin{figure}
\centerline{
  \includegraphics[width=1.15\columnwidth,angle=0] {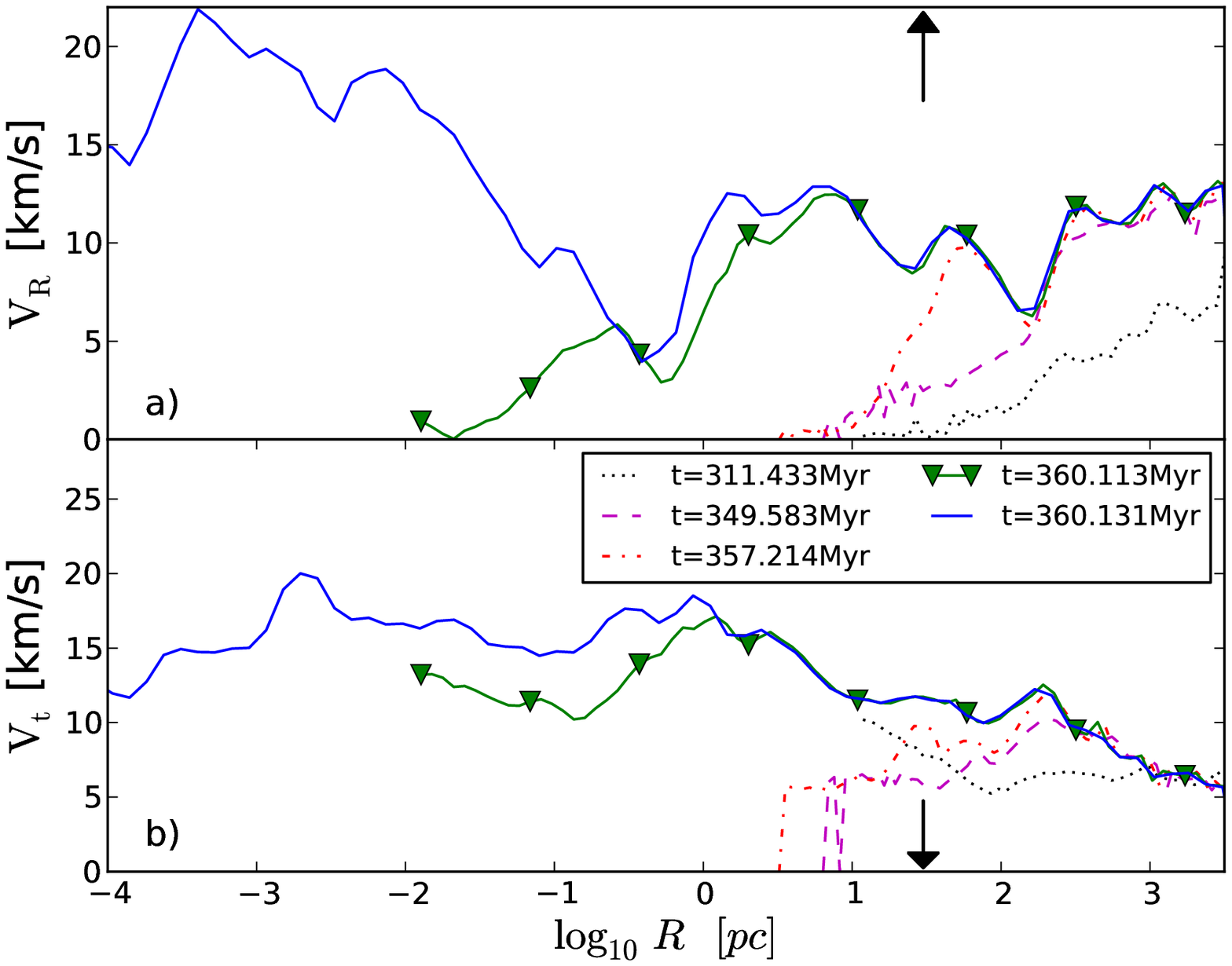}
}
\caption{Radial (upper) and tangential (bottom) velocities profiles for the gas at various times
within the DM halo.
The arrows indicate the location of $R_{\rm s}$
}
\label{fig:rtvel}
\end{figure}

The radial (infall) velocity and and $\dot{M}$ actually decrease from $R_{\rm vir}$ (800 pc) to $\sim 100$\,pc, 
where they have either global or local minima (see the upper frame of Figure\,\ref{fig:Mdot} and 
Figure\,\ref{fig:rtvel}). 
As the gas moves toward $R\sim 100$\,pc, it experiences an increase in density, $R^{-2.5-3}$, which is
offset by a sharper decrease in $v_{\rm R}$. To fully understand this behavior of $\dot M(R)$ and 
$v_{\rm R}(R)$, one should also note a simultaneous increase in the tangential velocity at the same radii
--- this means that the gas angular momentum becomes more significant. Inside
the minimum at $R\sim 100$\,pc, both the density and the radial velocity increase inward, which results
in the sharp increase of $\dot{M}$.

Inside $R_{\rm s}$, $\dot{M}$ reaches $\sim {\rm few}\,\Msun\,{\rm yr^{-1}}$, and 
stays approximately constant down to $R\sim 10^{-4}$\,pc. $R_{\rm s}$ is roughly 
the radius at which the gas density approaches the DM density, and the gas decouples from the background 
DM potential. The value of $\dot{M}$ can be estimated from the isothermal spherical gas collapse 
model, $\sim (M_{\rm gas}/M_{\rm tot}) v_{\rm ff}^3/G$, where the free-fall velocity, $v_{\rm ff}$, is a 
measure of the {\it local} 
gravitational potential, and $M_{\rm gas}/M_{\rm tot}$ is the gas-to-total mass ratio within $R$ 
\citep{Choi.etal:13}. The virial temperature inside 
$R_{\rm s}$ is $T_{\rm gas} \sim 8000$\,K, which gives the observed value of
$\dot{M}\sim {\rm few}\,\Msun\,{\rm yr^{-1}}$. Note that
we continue this simulation until the collapse has been established all the way to $R\sim 10^{-4}$\,pc,
the radius where we estimate that the optical depth for the radiation produced internally by the gas
will reach unity \citep[e.g.,][]{Choi.etal:13}. The isothermal density profile for the gas, therefore, has
been established between this radius and $R_{\rm s}$. 
When the collapse is allowed to proceed further,
we expect that the virial temperature of the decoupled gas will exceed the halo virial temperature
by a large factor (even if the gas density profile flattens), as the gas will determine the depth of the potential well. 
The actual temperature of the optically-thick collapsing gas, of course, depends on  
radiative transfer effects (and other cooling mechanisms), first for the bound-bound transitions and then for 
the continuum.

This rapid gas accretion is one of the key differences between the physical conditions during star 
formation and SMBH seed formation \citep{Begelman.etal:06,Begelman.Shlosman:09}. The cosmological 
simulation in this paper demonstrates that the gas collapse, which is facilitated through atomic cooling, 
reproduces the central runaway with high $\dot{M}$ and without significant fragmentation. Our results 
show that the gas collapse proceeds roughly in two stages, reflecting the shape of the
gravitational potential dominated by the DM initially and by the gas thereafter.

\subsection{Dynamics of the collapsing gas}
\label{sec:dyn}

\begin{figure*}
  \includegraphics[width=1.\textwidth,angle=0] {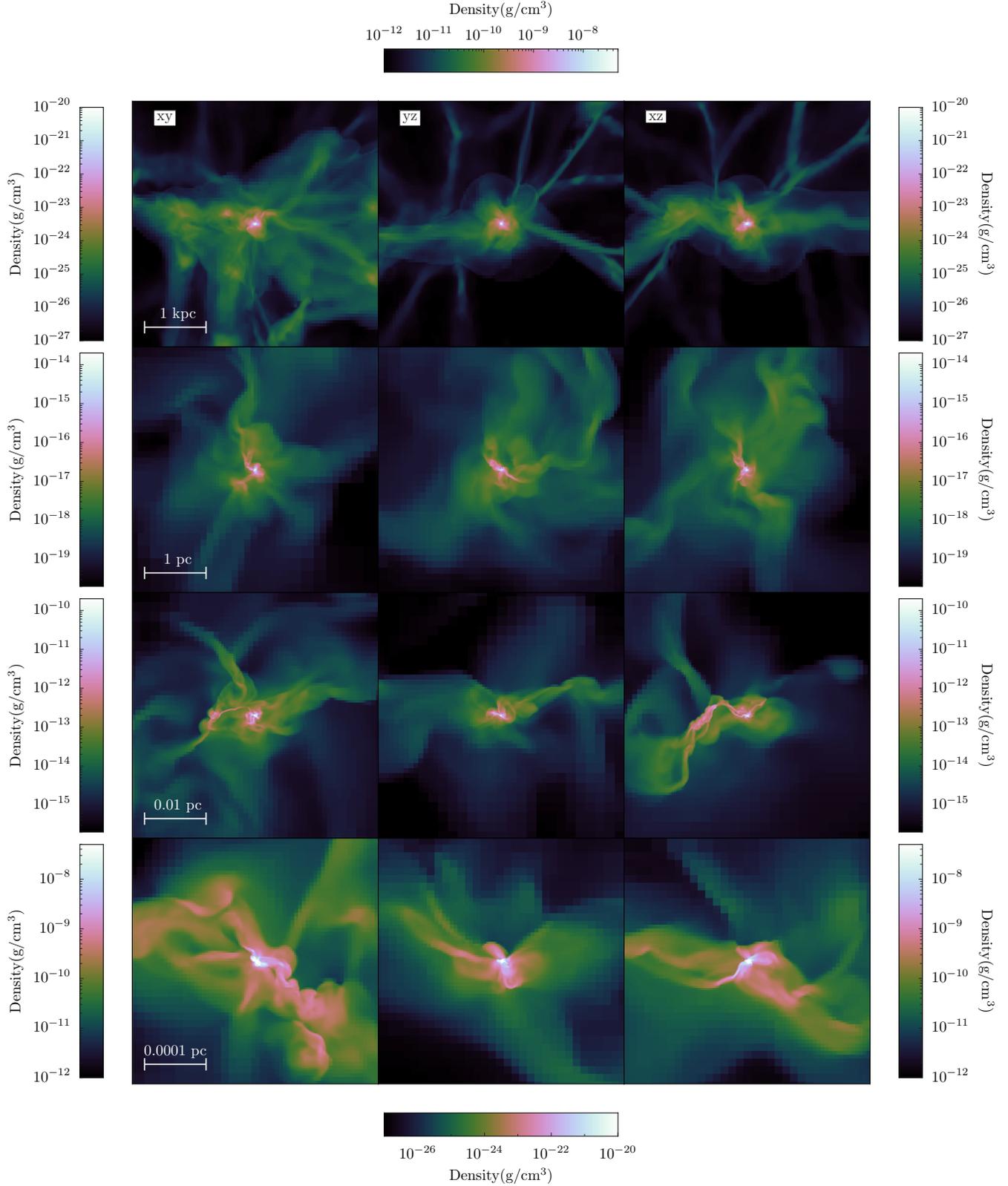}
  \caption{Gas density slices in three projections, on scales 1\,kpc$-10^{-4}$\,pc (in physical coordinates),
    of the runaway gas collapse centered on the highest gas density cell, at the end of the simulation, 
    $t\sim 360.13$\,Myr.
    The top row shows the targeted halo scale, and each subsequent row shows the gas
    with the scale zoomed by a factor of 1,000 or 100.
    Continuous gas collapse can be observed over seven decades in radius.
    Filamentary structures form at all radii, and the dominant elongated structure can be observed
    on scales $\ltorder 1$\,pc. No significant fragmentation has been observed.
  }
\label{fig:VelDen}
\end{figure*}

Evolution of radial and tangential velocities in the collapsing gas is shown in Figure~\ref{fig:rtvel}.
Note that the highest velocities are achieved inside the central pc, where the gas has decoupled from the 
background DM potential and increases its virial velocity above that of the DM. The tangential
velocity approaches $\sim 20\,{\rm km\,s^{-1}}$ and so does its radial counterpart
--- a clear indication that the gas has substantial rotational support, albeit sub-Keplerian, as it 
continues to collapse. We shall return to this issue again.

Figure~\ref{fig:VelDen} displays gas density slice maps at the end of the simulation, 
on four characteristic (physical) scales, from 1\,kpc down to $10^{-4}$\,pc,
on the $xy$, $xz$ and $yz$ planes. The filamentary gas distribution, which can be observed on the largest scales, 
has its origin in the DM distribution.  The top row exhibits 
the overall environment of the growing halo, on scales of $\sim 1$\,kpc.
Multiple filaments which fuel the accretion can be observed on this scale. 
The second row ($\sim 1$\,pc) shows the inner halo where the gas density becomes higher than 
the DM density. The third and last rows, $\sim 0.01-10^{-4}$\,pc, display regions fully dominated by the 
gas and very asymmetric in distribution and filamentary. the last row, at $\sim 200$\,AU, represents the 
environment of the expected optically thin-to-thick transition in the collapsing gas.

The continuity of the filamentary structure to small radii hints that its origin does not lie in shocks 
but rather that this is inherently a signature of a cosmological accretion flow before it virializes. Note that
we stop the simulation very early, but already at this time we see the non-axisymmetric gas response to
a non-axisymmetric background potential completely dominated by the DM in the top row frames. Already
in the second row of frames, the gas potential dominates, but the DM on larger scales can still provide the 
finite amplitude non-axisymmetric perturbation on smaller scales\footnote{In a non-axisymmetric density 
distribution, the material at larger radii exerts gravitational torques on the interior material.},
assisted by the non-axisymmetric distribution of the gas on all radii.
The gas density slices in the bottom row frames clearly show the presence of a non-axisymmetric feature centered 
on the highest gas density cell. The apparent presence of low-level Fourier harmonics is verified by the mode 
analysis of the gas density described below. 
These non-axisymmetric features play a central role in the transfer of angular momentum 
outward by gravitational torques on small scales --- a process that allows for the continuous gas collapse.

Despite the fact that the halo has only a small fraction of rotational support ($\lambda\sim 0.03$), as
does the gas, the centrifugal support of the gas would increase quickly if angular momentum were conserved 
during the collapse.
Without efficient transfer of angular momentum, the collapse would be 
halted when the angular momentum reached its Keplerian value. In the presence of the background DM, this 
corresponds to collapse by a factor of 10 in radius from the largest scales \citep[e.g.,][]{Shlosman:13}. As 
gas does not accumulate at 
any radius in our simulation (although it does slow down its radial motion at various radii), it is clear 
that it does not reach the centrifugal
barrier, and, therefore, its angular momentum is not conserved. That non-axisymmetric perturbations
can facilitate angular momentum transfer in steady state systems is well known --- these can be spiral arms 
\citep[e.g.,][]{Lynden-Bel.Kalnajs:72,Tremaine.Weinberg:84}, large-scale bars or a hierarchical bars-in-bars 
structure \citep[e.g.,][]{Shlosman.etal:89,Shlosman:05}. 

While spontaneous bar instability typically requires 
a few rotation periods to develop, bars and bars-in-bars can also be triggered by a finite amplitude
perturbation, e.g., such as provided by triaxial DM halos \citep[][]{Shlosman:11}.  
Strong gravitational torques which accompany finite amplitude perturbations can transfer angular momentum
on the short dynamical timescale encountered in direct collapse, and are more efficient as their rise time
is negligible. Such torques can follow from the low-$m$ Fourier 
non-axisymmetric modes, like $m=1$ and $m=2$, which can be associated with the displacement of the center of mass 
of the gas with respect to the DM, and/or the development of nested gaseous bars 
\citep[][]{Shlosman.etal:89,Shlosman.etal:90,
Englmaier.Shlosman:04,Begelman.Shlosman:09}. We, therefore, analyze the prevailing non-axisymmetric modes 
arising in our simulations (e.g., Figure~\ref{fig:VelDen}).

\begin{figure}
  \includegraphics[width=1.0\columnwidth,angle=0]{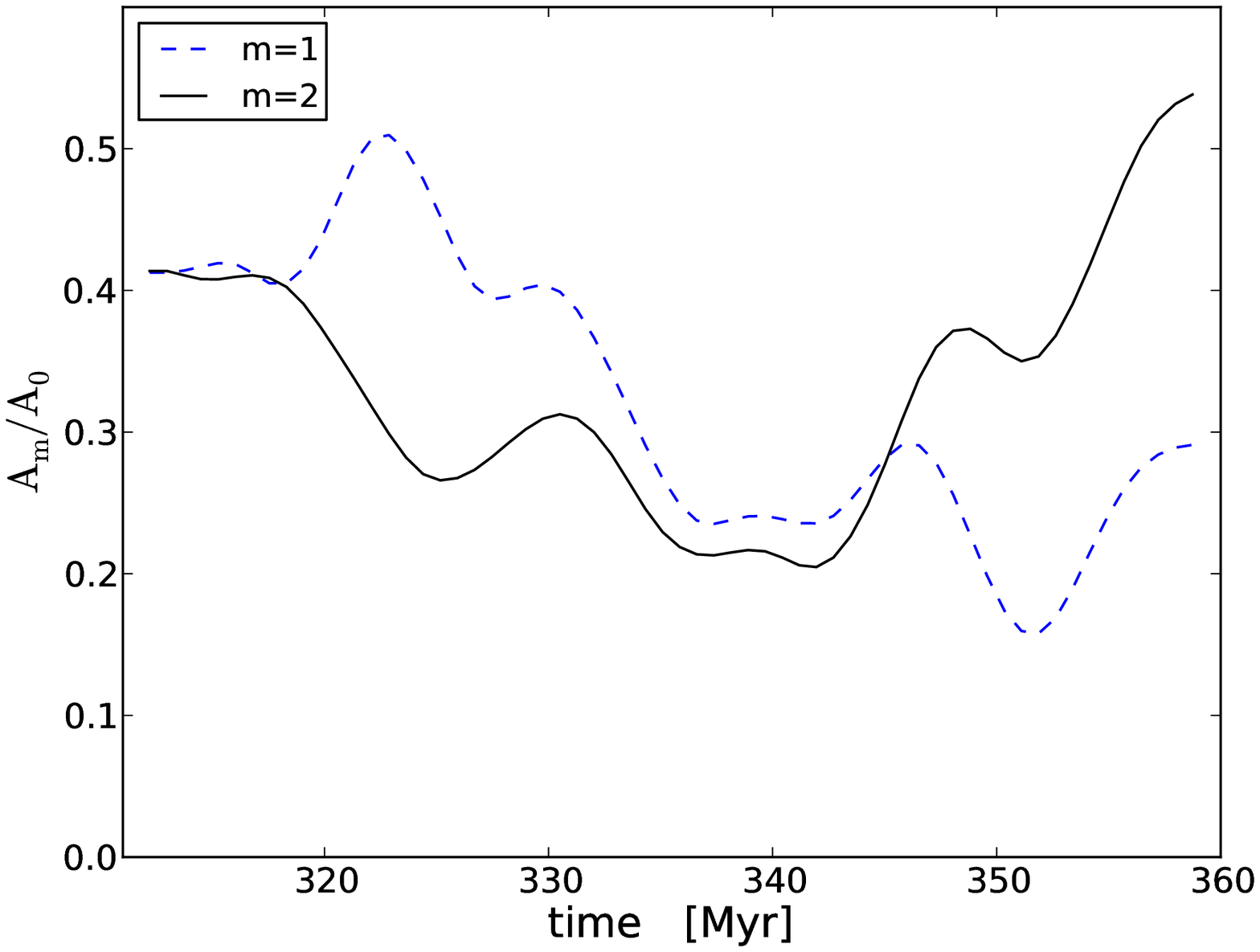}
  \includegraphics[width=1.0\columnwidth,angle=0]{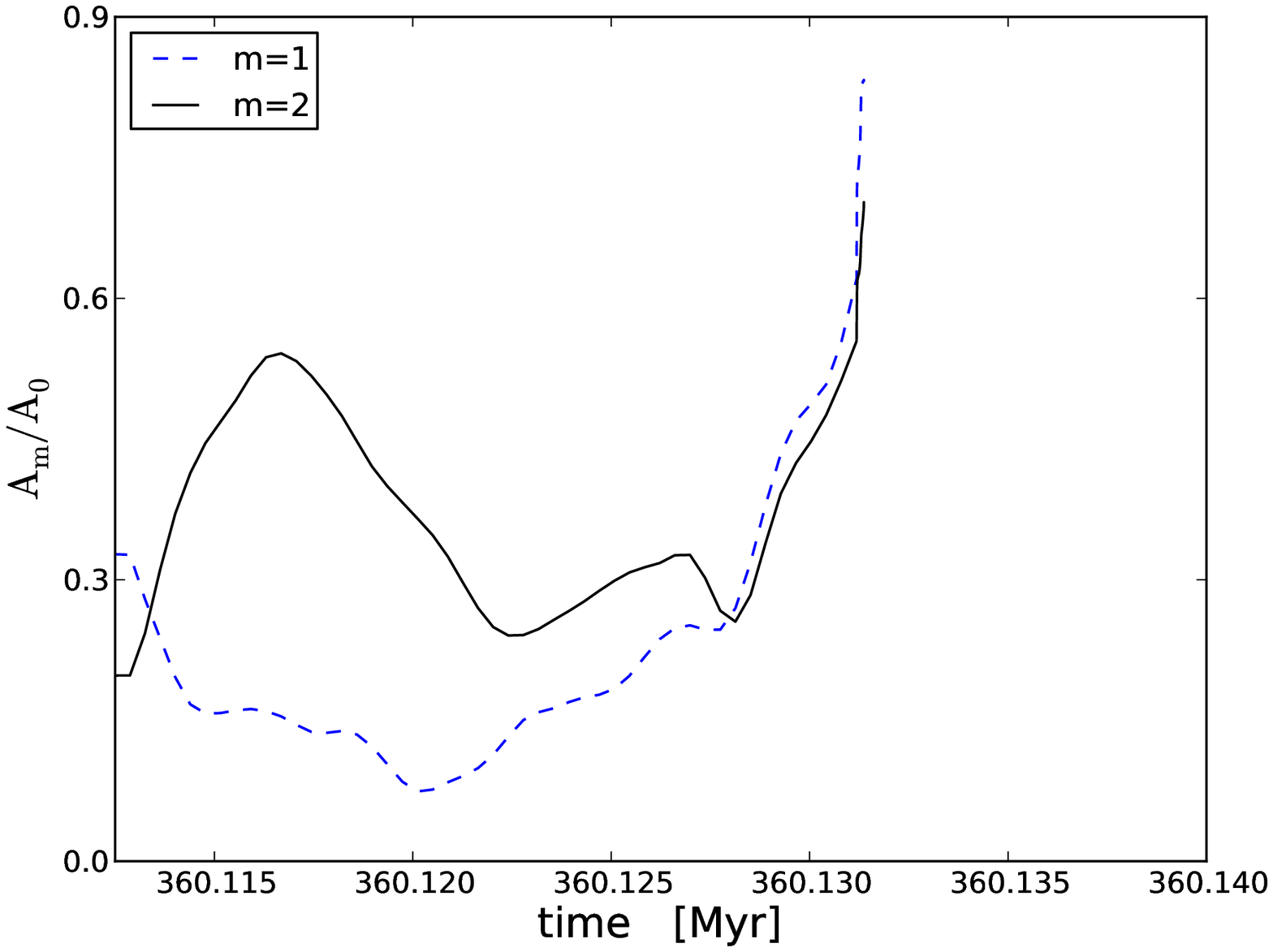}
  \caption{Evolution of the Fourier modes $m=1$, and 2 normalized by the $m=0$ mode amplitude.
    The top panel is measured within the cylindrical annulus defined by $10\leq r\leq 50$\,pc and 
    thickness $\Delta z = 5$\,pc with respect to the $xy$-plane. The time frames are chosen to
    capture mode evolution during the large scale collapse, including the central runaway.
    The bottom panel is measured within an annulus of $0.01\leq r\leq 0.1$\,pc and $\Delta z = 0.01$\,pc with 
    respect to the same plane. The time frames include the central runaway only. The plot is smoothed to reduce 
    the data noise.
}
\label{fig:mode}
\end{figure}

Although Figure~\ref{fig:VelDen} clearly indicates the existence of non-axisymmetric density features in the 
inner halo, it is not clear what mode dominates the structure. We, therefore, perform a Fourier analysis
for two density modes, $m=1$ and 2 --- the fastest growing non-axisymmetric modes 
(see e.g., \citet[][]{Long.etal:14} for technical details). The $m=1$ mode requires 
displacement of the center of mass of the 
system. Its presence is evident in the fact that the densest cell in the simulation separates from the center 
of mass of the collapsing gas,
measured within a sphere with a radius of 0.08\,pc.
The $m=2$ mode is a barlike mode and usually plays the dominant role in angular momentum transfer.
Figure~\ref{fig:mode} shows the evolution of the Fourier amplitudes of $m=1$ and 2 modes normalized by the 
amplitude of the $m=0$ mode. The mode evolution shown starts after the time when atomic
cooling has been triggered in the halo, $\sim 310$\,Myr.  We display the mode evolution
only for one plane, but it is representative of the overall behavior. The DM halo shape is that of triaxial 
ellipsoid and exerts
gravitational torques on the gas. The latter responds to this finite amplitude perturbation in a nonlinear 
fashion and develops $m=1$, 2 and higher modes. 

The top panel in Figure~\ref{fig:mode} shows the mode evolution for the cylindrical annulus defined
by the minimal and maximal radii of $r_{\rm min}=10$\,pc and $r_{\rm max}=50$\,pc in the $xy$-plane, with
the vertical thickness of the slice $\Delta z=5$\,pc.  This region is dominated by the DM, and the behavior of
both modes is similar to that of the isolated halo studied by \citet[][]{Choi.etal:13}, although the amplitude of 
both modes is higher. Before the gas decouples from the DM, the $m=1$
mode dominates over the $m=2$ mode. However, around the start of the central runaway, $\sim 355$\,Myr,
the $m=2$ mode starts to dominate over $m=1$.
The bottom panel shows mode evolution for a smaller
annulus defined by $r_{\rm min}=0.01$\,pc,  $r_{\rm max}=0.1$\,pc and $\Delta z=0.01$\,pc, respectively. This
scale is dominated by the gas. It shows only the last stage of the central runaway, where $m=2$ dominates
over the $m=1$, until the very last moment where both of them  increase dramatically, reflecting the
formation of the elongated structure --- the nonlinear response of the very central gas.
Comparing the large and small scale evolution of these modes, we note that last rise of the $m=1$ mode is important 
only at the
very center. It has a much smaller amplitude on the larger scale, indicating that the cause for its rise lies in the
off-center motion on scales $\ltorder 0.1$\,pc.

\subsection{Collapse and angular momentum}
\label{sec:angmom}

The angular momentum axis of the collapsing gas varies with $R$ --- an
effect discussed by \citet[][see their Fig.\,19]{Romano-Diaz.etal:09} for the growth of DM halos in the
cosmological context. It reflects the variability in the angular momentum axis of the continuous gas inflow.
Since the observer's frame is fixed in its position, the gas moving to smaller radii is replaced by gas 
inflow from larger radii. This `replacement' gas has typically a different orientation of the angular momentum 
axis, and can in principle drive variability  of the Fourier density mode amplitude.  

\begin{figure*}
\includegraphics[width=0.49\textwidth,angle=0]{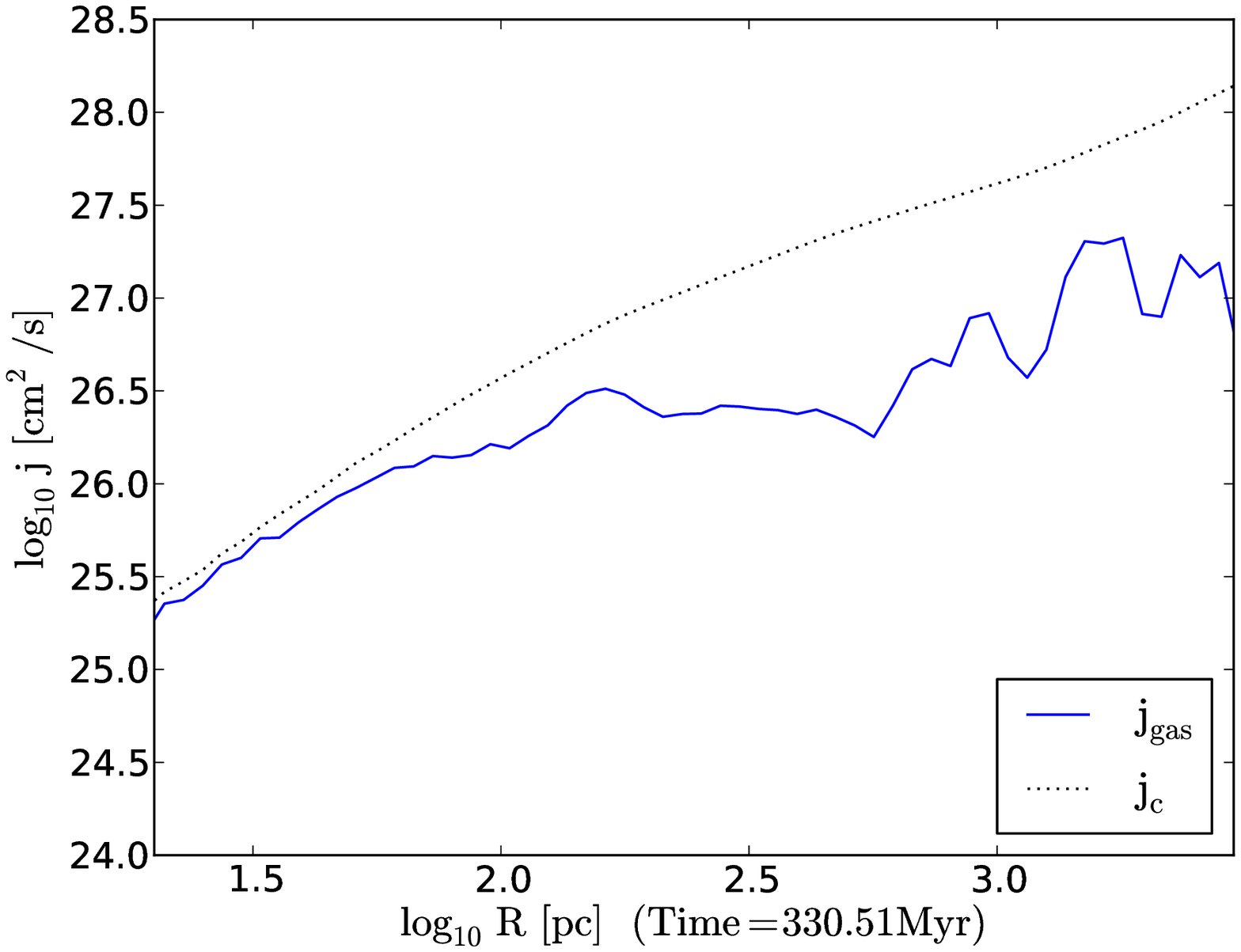}
\includegraphics[width=0.49\textwidth,angle=0]{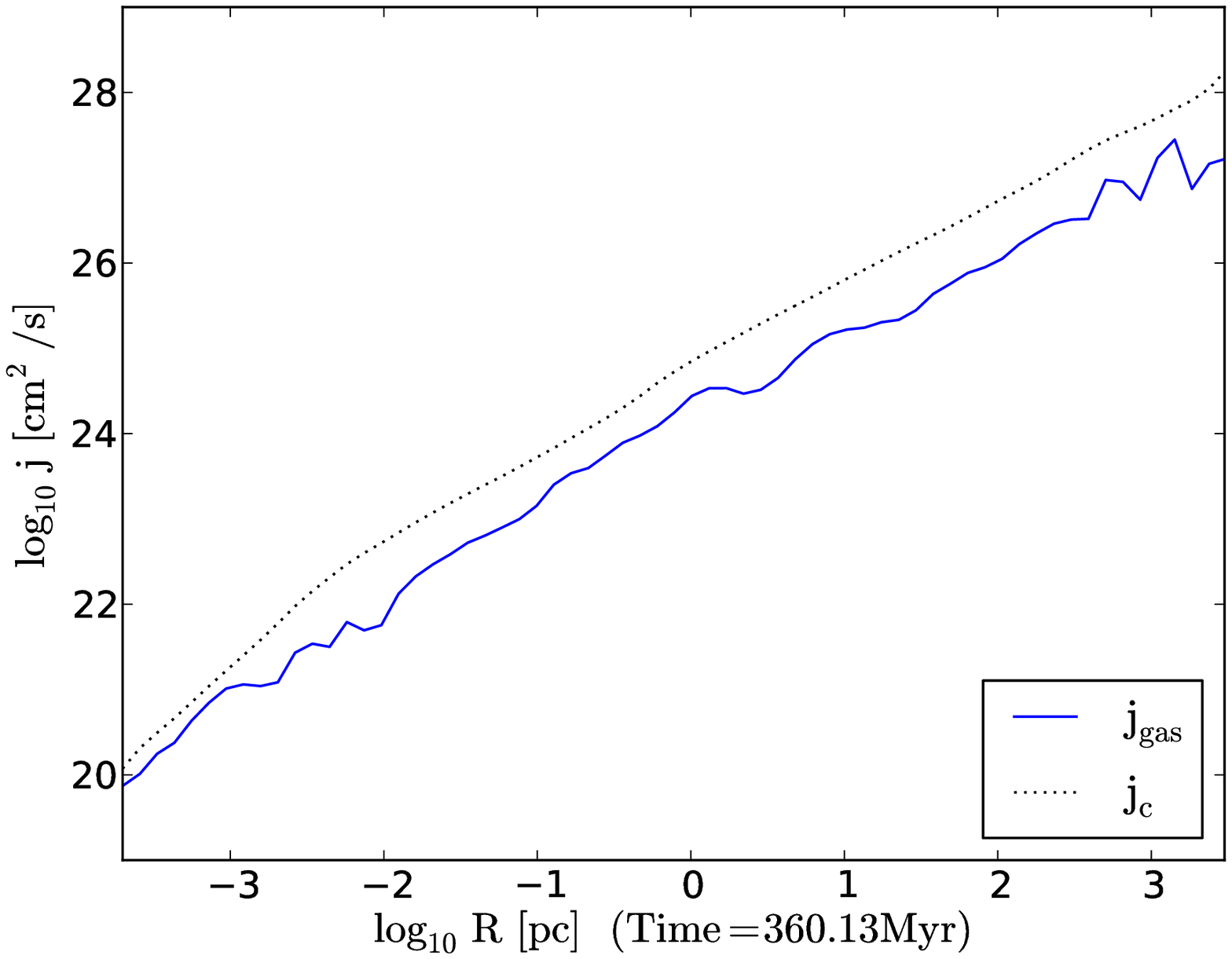}
 \caption{Specific angular momentum profiles, $j_{\rm gas}(r)$, for the gas during the central runaway collapse 
    at two snapshots, $t=330.51$\,Myr (first stage of the collapse) and 360.13\,Myr (second stage, end of the 
    simulation), shown by 
    the blue solid line. The dotted line represents the specific angular momentum for the circular rotation, 
    $j_{\rm c}(r)$. Note that $j_{\rm gas}$ is close to $j_{\rm c}$ at smaller radii in each snapshot.
}
\label{fig:jevol}
\end{figure*}

Figure~\ref{fig:jevol} provides two snapshots of the specific angular momentum profile of the gas, 
$j_{\rm gas}(R)$. These profiles are superposed on the circular specific angular momentum $j_{\rm c}(R)$.
At each radius, $j_{\rm c}$ is the maximum possible angular momentum allowed for the bound gas at a fixed 
energy.
The total mass included within a sphere of radius $R$ is about linear with $R$ in the halo, the circular 
velocity is about constant, and $j_{\rm c}\sim R$. $j_{\rm gas}(R)$ is typically smaller than
$j_{\rm c}$, by a factor that varies with $R$ and with time.  
This means that the collapse is typically not prohibited by the angular momentum barrier, for a considerable 
range in radii. At smallest radii, it lingers not far from the centrifugal barrier, at each time. This 
can happen only if there is a constant flow of angular momentum away from the gas as it moves inward.

As the DM distribution is triaxial, it exerts gravitational torques on the gas at all radii, as we noted
above, but its contribution is gradually washed out in the region where the gas dominates. The gas is losing 
its angular momentum at a rate of $dj_{\rm gas}/dt\sim \tau$, where $\tau$ is 
the torque per unit mass. It depends on the offset in the position angle between the gas and DM, i.e.,
on the asymmetric parts of the DM and gas density distributions, and the asymmetric part
of the background gravitational potential \citep[e.g.,][]{Berentzen.etal:07}. The angular momentum,
therefore, can flow from the inner gas to the outer gas, and from the gas to the DM. 

The efficiency of angular momentum extraction by gravitational torques can be measured
by $\eta\equiv 1-j_{\rm gas}/j_{\rm c}$, i.e., by the `separation' between the centrifugal barrier
and the actual angular momentum in the gas at each $R$. The right frame of Figure~\ref{fig:jevol} 
displays the typical situation in the later stages of the gravitational collapse. The efficiency $\eta$
is about constant for a wide range in radii. Note, that if $j_{\rm gas}$ moved away from $j_{\rm c}$
as the gas moves in, $\eta$ would increase and the residual $j_{\rm gas}$ would be less important
dynamically. On the other hand, if $j_{\rm gas}$ decreased more slowly than $j_{\rm c}$, the angular
momentum would become more important. The fact that $\eta\approx $\,const., means that there exists a tight
balance between the characteristic timescale of $j$ loss by the gas and the inflow timescale.  

The gas distribution is substantially asymmetric on all scales (Figure~\ref{fig:jevol}) and is
subject to strong gravitational torques.  The bottom panel of Figure~\ref{fig:mode} confirms that the
$m=1$ and 2 modes are the key contributors to the transfer of the angular momentum.
Figure~\ref{fig:mode} and scrupulous inspection of the full
evolutionary feature of the angular momentum profile at various times confirm that the angular momentum 
transfer happens multiple times on multiple spatial scales.
The figure rotation of the DM halo can be neglected, as it tumbles extremely slowly 
\citep[][]{Romano-Diaz.etal:09}. The outer torques are dominated by the DM distribution, while at smaller 
radii, it is the nonaxisymmetric shape of the gas distribution that provides the torques. 

Unlike in the case of collapse from idealized conditions within an isolated halo, the cosmological collapse,
expectedly, does not show the formation of a disk, because rotational support never becomes high enough
and because the vector of the specific angular momentum exhibits temporal and spatial variability.

\begin{figure*}
\centerline{
  \includegraphics[width=1.\textwidth,angle=0] {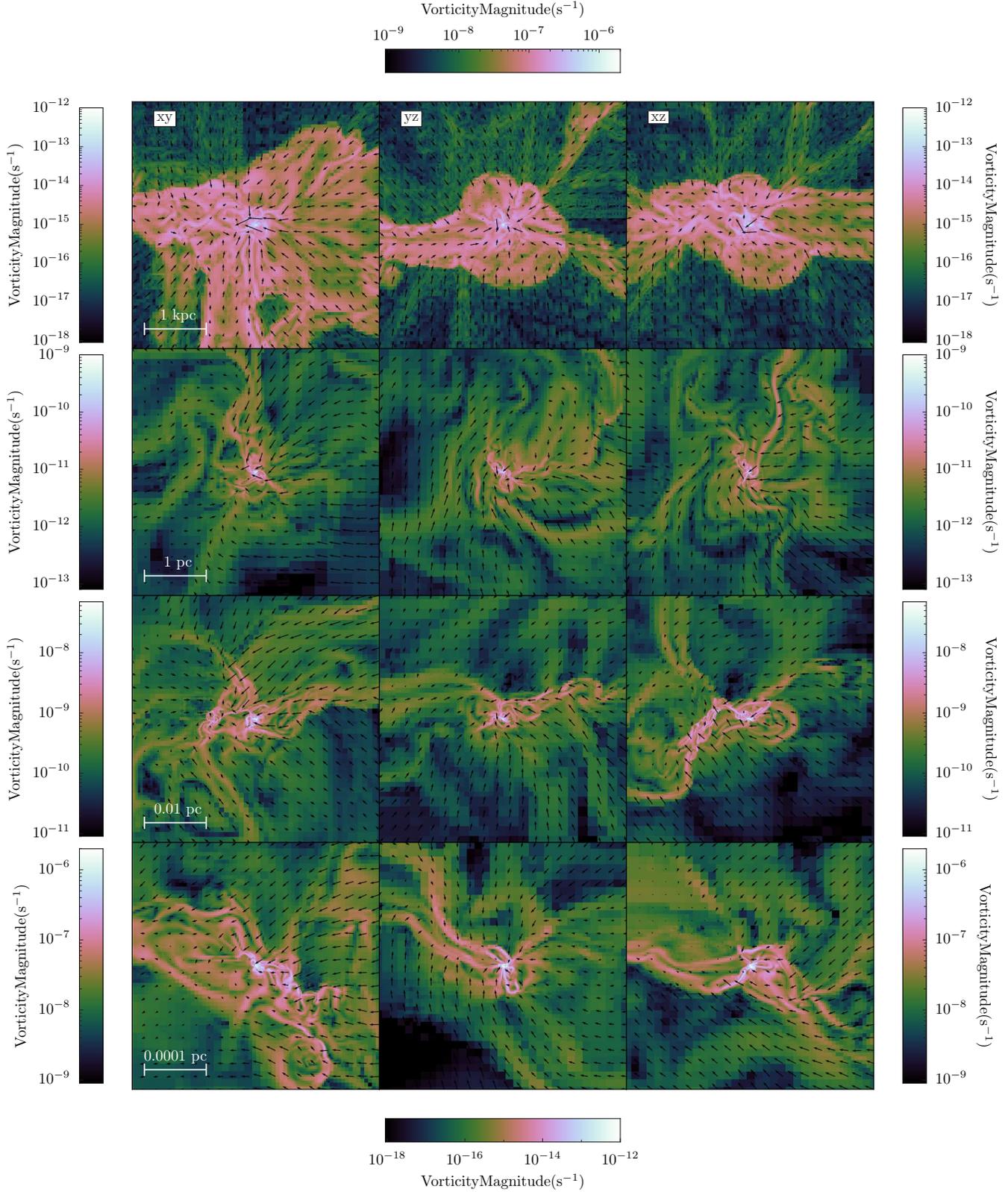}}
  \caption{Vorticity magnitude slice with projected velocity arrow of the runaway collapse, in physical
  coordinates, at the end of the simulation at $t\sim 360.13$\,Myr.
    The color represents the vorticity magnitude and the arrows represent the projected gas velocity.
    The length of the arrow provides the relative magnitude of the velocity.
    The existence of vorticity indicates the presence of a turbulent gas motion.
    The vorticity magnitude increases on a smaller scale.
    The velocity arrows show that the gas motion on the large scale is mostly radial, while the rotation 
    has developed on a smaller scale.
  }
\label{fig:Vorticity}
\end{figure*}

\subsection{Collapse and turbulence}
\label{sec:turb}

Figure~\ref{fig:Vorticity} displays slices of the collapsing gas with the gas velocities and the degree of
turbulence given by the magnitude of the vorticity, defined by ${\rm\bf w} = {\rm\bf \nabla\times v}$, where 
${\rm\bf v}$ is the velocity field.
The evolutional time, spatial scales, and viewing angles are the same as in Figure~\ref{fig:VelDen}.
The arrows represent the direction of gas flow and their sizes give the 
relative flow speeds. The top panels exhibit the gas motion on the scale of the halo, 1\,kpc.  
The middle panels show the gas flow on scales where it decouples from the DM ---
rotational motions are clearly visible and so some degree of rotational support is present. Similar 
phenomena can be observed on the smaller scales (bottom panels).

Figure~\ref{fig:Vorticity} demonstrates the degree of turbulence.
The  collapsing halo gas develops the supersonic turbulent motions 
\citep[e.g.,][]{Wise.etal:08,Begelman.Shlosman:09,Regan.Haehnelt:09}, which suppress fragmentation 
\citep{Begelman.Shlosman:09,Choi.etal:13}. 
The supersonic turbulence works to both damp and trigger fragmentation.
The damping is provided via turbulent pressure, above the level of the thermal pressure, thus acting against
the self-gravity. On the other hand, shocks associated with supersonic turbulent motions induce fragmentation
\citep[e.g.,][]{Krumholz.McKee:05}.
The fragments, however, must collapse before the passage of the next shock front, which otherwise will
destroy the fragments. We also note that fragmentation in the DM-dominated phase is suppressed by the
DM background, because its action dilutes the gas self-gravity, and, therefore, increases its Jeans mass ---
the gas cannot collapse until its density surpasses that of the background DM \citep[][]{Choi.etal:13}. 

In Figure~\ref{fig:Vorticity}, the smaller scales exhibit larger vorticity magnitude than 
larger scales (note the shifting colors in the color palettes). This implies that the turbulent motions 
increase and will continue to increase with the gas collapse, as the potential well generated by the gas 
deepens. This is consistent with Figure~\ref{fig:rtvel}: $v_{\rm R}$ and $v_{\rm t}$ increase  
at smaller $R$.

\begin{figure}
\centerline{
   \includegraphics[width=0.55\textwidth,angle=0]{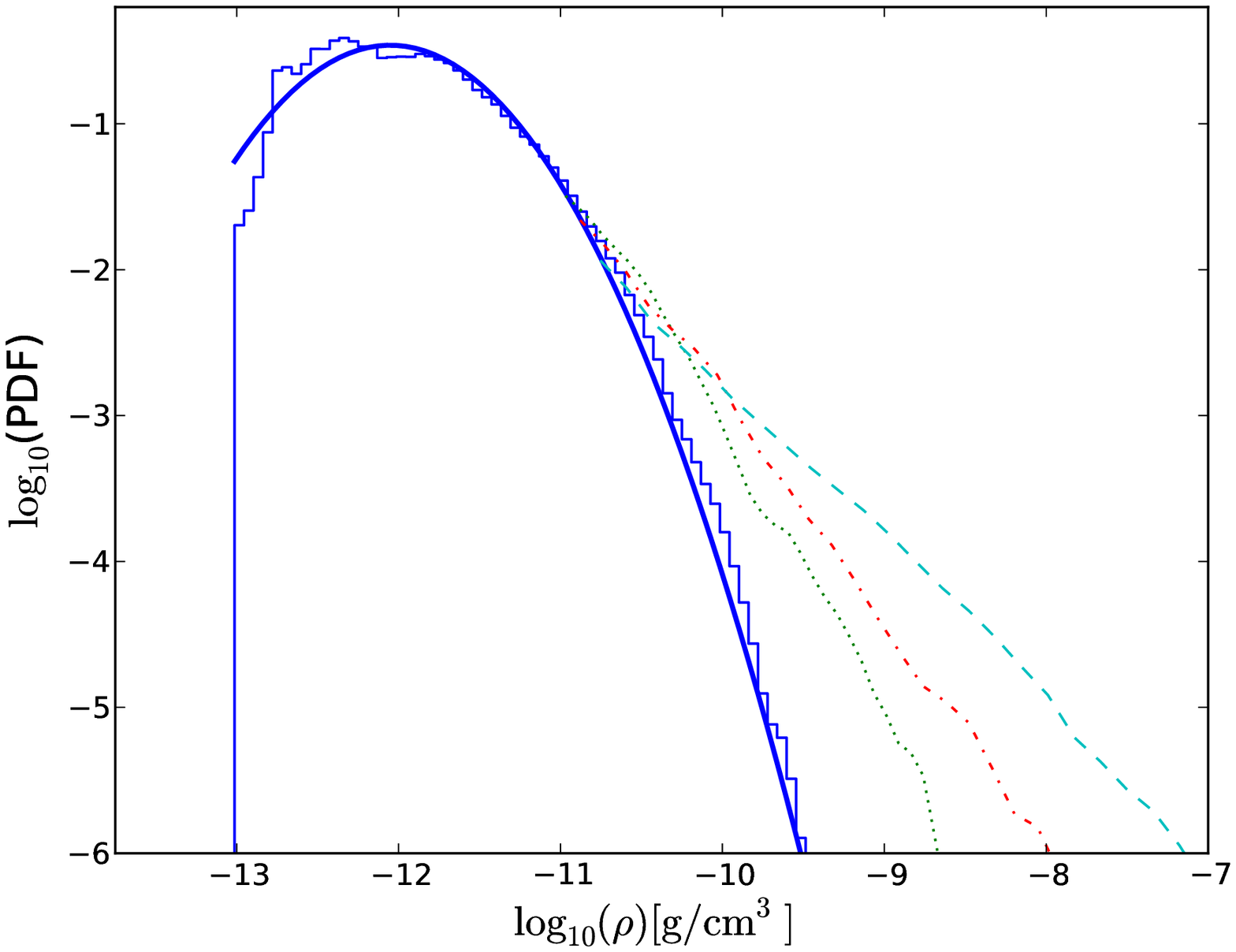}}
   \caption{Evolution of the volume-averaged, gas density PDF as a function of $\log_{10} \rho$ measured at the end 
     of the simulation, at $t\sim 360.13$\,Myr, and sampled with $\gtorder 10^6$ AMR cells at the fixed
     refinement level (see text). The sampling shows the PDFs of the central sphere of radius 200\,AU as blue histogram.
     Shown also are the lognormal fit (thick blue solide line)
     presented in Equation~\ref{eq:pdf} with $\sigma\sim 1.17\Mach$. The average density for the sampled sphere is 
     $\overline{\rho} = 1.76\times 10^{-12}\, {\rm g\,cm^{-3}}$ (for the blue histogram).
     It also shows the continuous evolution of the high density side, $\rho > 10^{-10} {\rm g\,cm^{-3}}$, of the blue histogram
     at consecutive times separated by $\sim 10$\,yrs as green dotted, red dot-dashed, and cyan dashed lines.
     The evolution shows that the lines have reached the slope of $\sim -1.18$ in the power law tail.
     The collapsing gas has been sampled at the resolution of $\sim$0.7\,AU and the density fluctuations extend over 
     7 decades. The refinement levels for these figures is kept at 28, in order to compare with identical conditions
     in Figure 15 of \citet{Choi.etal:13}.
   }
\label{fig:denPDF4}
\end{figure}

Figures~\ref{fig:VelDen} and \ref{fig:Vorticity} show that no major gas fragmentation occurs.
A central density maximum is well-identified at all times.
Although the density slices do not display major fragmentation, multiple shocks are present as the analysis
shows. These shocks result from the supersonic turbulent motion of the collapsing halo gas.
The existence of the shocks without major fragmentation suggests that incipient fragments in the collapsing halo gas 
are destroyed by the next incoming shock, before the fragmentation proceeds into a strongly nonlinear regime. 
If fragmentation occured, it would deplete the available gas supply to the center and would disturb the
developing low-$m$ Fourier density modes which facilitate the angular momentum transfer. More quantitative analysis 
of this is given elsewhere.
 
Next, we analyze the turbulent motions using the gas density probability distribution 
function (PDF), following \citet{Choi.etal:13}.
Studies of supersonic turbulence find that such a PDF has a lognormal  shape
\citep[e.g.,][]{Vazquez-Semadeni:94,Padoan:95,Scalo.etal:98,Ostriker.etal:99,Padoan.Nordlund:02,Krumholz.McKee:05,
Choi.etal:13,Federrath:13}:
\begin{eqnarray}
p(x) = \frac{1}{(2 \pi \sigma^{2}_{\rm p})^{0.5}} \frac{1}{x} \exp \biggl[ -\frac{(\ln x - \overline{\ln x})^2}
   {2\sigma^{2}_{\rm p}}\biggr],
\label{eq:pdf}
\end{eqnarray}
where the distribution mean $\overline{\ln x} = -0.5 \sigma^{2}_{\rm p}$, $x \equiv \rho/\rho_{0}$ ($\rho_0$ being 
the mean density), and its dispersion is $\sigma^{2}_{\rm p} \sim [\ln(1 + 3\Mach^{2}/4)]$. However, in the presence
of a self-gravitating gas, the lognormal distribution has been found to develop a power law tail 
\citep[e.g.,][]{Vazquez-Semadeni:94,Kritsuk.etal:11,Choi.etal:13}.

Our simulation has been stopped when the collapse reaches the required scale of $\sim 10^{-4}$\,pc, which happens
when the refinement level has reached 35. In order to compare directly with the simulations of \citet[][]{Choi.etal:13},
we have restarted the simulation when it reaches the comparable refinement level of 28. 
We sample the gas density cells within a distance of 200\,AU from the center of the collapse with $\gtorder 10^6$ 
cells, and fit the PDF (by least squares) to a pure lognormal distribution with $\sigma\sim 1.17\Mach$ (blue histogram) 
from 
Equation~\ref{eq:pdf}. We have also examined several different sampling scales and locations and obtained similar 
results. Figure~\ref{fig:denPDF4} confirms that the obtained blue histogram PDF is nicely fit with the lognormal 
distribution, without any visible power law tail. We, therefore, continue the simulation for additional time at a 
fixed refinement level. Figure~\ref{fig:denPDF4} shows the evolution of the density PDF and one can clearly observe 
the formation of a progressively shallow power law tail. The terminal slope (cyan dashed line) is about $-1.18$.
As the power law tail appears at high densities, its origin must be related to the onset of self-gravity as the
gas starts to pile up in the center. However, the power law slope  attained in this simulation is 
still evolving, and its terminal value most probably would steepen as the collapse proceeds.

\section{Discussion}
\label{sec:discuss}

In this paper, we have investigated some aspects of direct collapse that can lead to the formation of SMBH seeds 
at high redshifts. We have used the cosmological zoom-in initial conditions and an Eulerian AMR code to resolve the  
gas and DM dynamics inside and in the vicinity of the targeted DM halo. We have shown that the gas atomic cooling in 
the DM halo will trigger central runaway collapse without significant fragmentation.
The central runaway can be divided into two stages: outer collapse in the DM-dominated potential, where the density
flattens gradually from a log slope of $-3$ to $-2$ at a few pc from the center, i.e., the NFW profile, and inner
collapse at smaller $R$. 
The characteristic radius where the second stage is triggered corresponds roughly to the NFW scale radius $R_{\rm s}$,
where the background DM density gradually becomes shallower and reaches the (log) slope of $-2$.
The associated mass 
accretion rate reaches a few\,$\Msun\,{yr^{-1}}$ at this radius. The second stage of the collapse represents
the gas-dominated region, where the gas decouples from the DM background potential and continues its collapse.
As the physical conditions in the central regions require the introduction of radiative transfer on-the-fly, due to 
the buildup of optical depth in the hydrogen lines, we have terminated
the collapse in the early stage when it reaches the required resolution scale of $10^{-4}$\,pc.

We confirm the \citet{Choi.etal:13} finding that the low-$m$ Fourier non-axisymmetric modes are responsible for the 
angular 
momentum flow outward which results in overcoming the angular momentum barrier at various radii. The specifics
of direct collapse in the cosmological context involves the variability of the angular momentum vector as a function 
of radius and time. Our analysis of the mode and angular momentum evolution demonstrates that 
gravitational
torques clearly dominate over the hydrodynamical torques over a wide dynamic range during the collapse.
Efficient loss of angular momentum allows for continuous collapse over 7 decades in radius, compared to
one decade if the angular momentum were conserved in the gas.
We also demonstrate that the collapsing flow develops supersonic turbulent motion.
The degree of the turbulence increases as the collapse proceeds to small radii, and
the supersonic turbulent motion suppresses gas fragmentation.

\citet[][]{Latif.etal:13a} have argued that turbulent eddies will be under-resolved at small $R$, and have
added the turbulent driving on all scales. However, we achieve the highest resolution at the smallest scales
and have tested both the rms velocities there, which appear to be supersonic, and the density PDF, which extends 
over 7 decades in density. We find
no indication that the turbulence is unresolved there. One can compare the final state of the simulation
achieved with and without turbulent driving. Indeed, the driving leads to a more regular rotation on small
scales, but one questions whether this is realistic. This issues must be studied further. 

Simulations of direct collapse in the cosmological context have been also performed by \citet[][]{Prieto.etal:13}, 
but their
spatial resolution has been limited to $\sim 1\%$ of the halo virial radius. This is insufficient to resolve
the second stage of the gravitational collapse, where the gas decouples from the DM background.

Additional processes are known to help to suppress fragmentation in the collapsing gas. The most important process is 
the dissociation of H$_2$ by UV background continuum from external sources.
It produces the temperature floor and maintains isothermality with $T\sim 10^4$\,K.
We assume that the UV background radiation prevents H$_2$ formation.
There are several recent studies arguing that this is a plausible case, when the Lyman-Werner background continuum
is present at sufficient levels \citep[e.g.,][]{Omukai:01,Shang.etal:10,Regan.etal:14}.
This radiation background can in principle constrain the population of massive SMBH seeds at high-$z$.

Another issue is related to the shrinking numerical timestep which allows us to follow the evolution of the
central regions but basically freezes the outer regions. A possible solution to circumvent the small timestep
is to impose a sink particle mechanism for a given resolution level 
\citep[e.g.][]{Latif.etal:13b}. In this case, the spatial resolution is limited by the scale imposed for the sink 
particles. The simulation can continue further and one can follow the long-term evolution of the direct collapse.
This approximation can provide some details of the evolution of the central objects (Shlosman, Choi, Begelman \& Nagamine 2015 in preparation).

Given that we have confirmed that the central runaway collapse can proceed without significant fragmentation, we
conclude that high-$z$ SMBH seeds can form, at least in principle, through a process of direct collapse within a 
DM halo.  SMBH seeds formed in this way, while relatively massive, are not expected to follow the low-redshift 
M-$\sigma$ relation. This suggests that the co-evolution of massive SMBH seeds and their host protogalaxies may 
follow a very different evolutionary path, particularly at high-$z$,  than their later counterparts. 

\section*{Acknowledgments}
We are grateful to Kentaro Nagamine, Long Do Cao and Yang Luo for many interesting and helpful discussions, and
thank the ENZO and YT support team. All analysis has been conducted using YT (\citet{yt},
http://yt-project.org/).  
J.H.C acknowledges support from NASA ATP NNX11AE09G, NSF AST-1009799, and Caltech/JPL SURP Project No. 1515294.
through The University of Texas at Austin (P.I. Paul Shapiro).
I.S. acknowledges support from NSF grant AST-0807760, from HST/STScI grant
AR-12639.01-A, and from International Joint Research Promotion Program at Osaka University. M.C.B. acknowledges 
support from the NSF under AST-0907872. Support for HST/STScI AR-12639.01-A was
provided by NASA through a grant from the STScI, which is operated by the AURA, Inc., under NASA contract NAS5-26555.


\label{lastpage}
\end{document}